\documentclass[10pt,twocolumn,twoside]{IEEEtran}
\IEEEoverridecommandlockouts
\usepackage{caption}
\usepackage{subfigure}
\usepackage{cite}
\usepackage{amsmath,amssymb,amsfonts}
\usepackage{algorithmic}
\usepackage{graphicx}
\usepackage{textcomp}
\usepackage{xcolor}
\usepackage{cite}
\usepackage{amsmath,amssymb,amsfonts}
\usepackage{algorithmic}
\usepackage{graphicx}
\usepackage{textcomp}
\usepackage{xcolor}
\usepackage{algorithm}
\usepackage{comment}
\usepackage{setspace}
\usepackage{color}
\usepackage{booktabs}
\usepackage{multirow}
\usepackage{epstopdf}
\usepackage{amsmath}

\def\BibTeX{{\rm B\kern-.05em{\sc i\kern-.025em b}\kern-.08em
    T\kern-.1667em\lower.7ex\hbox{E}\kern-.125emX}}
\begin{document}

\title{A Generative Learning Approach for Spatio-temporal Modeling in Connected Vehicular Network}
\author{\IEEEauthorblockN{Rong~Xia\IEEEauthorrefmark{1}, Yong~Xiao\IEEEauthorrefmark{1},  Yingyu~Li\IEEEauthorrefmark{1}, Marwan Krunz\IEEEauthorrefmark{2}, Dusit~Niyato \IEEEauthorrefmark{3} \\
\IEEEauthorblockA{\IEEEauthorrefmark{1}School of Electronic Information and Communications, Huazhong Univ. of Science \& Technology, China}\\
\IEEEauthorblockA{\IEEEauthorrefmark{2} Department of Electrical and Computer Engineering, University of Arizona, AZ}\\
\IEEEauthorblockA{\IEEEauthorrefmark{3} School of Computer Science and Engineering, Nanyang Technological University, Singapore}
}}
\maketitle

\begin{abstract}
Spatio-temporal modeling of wireless access latency is of great importance for connected-vehicular systems. The quality of the molded results rely heavily on the number and quality of samples which can vary
significantly due to the sensor deployment density as well as traffic volume and density. This paper proposes LaMI (Latency Model Inpainting), a novel framework to generate a comprehensive spatio-temporal of wireless access latency of a connected vehicles across a wide geographical area. LaMI adopts the idea from image inpainting and synthesizing and can reconstruct the missing latency samples by a two-step procedure. In particular, it first discovers the spatial correlation between samples collected in various regions using a patching-based approach and then feeds the original and highly correlated samples into a Variational Autoencoder (VAE), a deep generative model, to create latency samples with similar probability distribution with the original samples. Finally, LaMI establishes the empirical PDF of latency performance and maps the PDFs into the confidence levels of different vehicular service requirements. Extensive performance evaluation has been conducted using the real traces collected in a commercial LTE network in a university campus. Simulation results show that our proposed model can significantly improve the accuracy of latency modeling especially compared to existing popular solutions such as interpolation and nearest neighbor-based methods.
\end{abstract}

\vspace{-0.2in}
\section{Introduction}
Fog computing-supported smart vehicular system is recently gaining strong momentum due to its capability of offering computationally intensive services that exceed the capacity of in-vehicle computers. According to \cite{Market}, by 2020, there will be 153.6 billion in total connected vehicles on the road globally, creating a market of over 122.51 billion US dollars. Being recognized as the new way to create business opportunities and increase revenues, major telecommunication mobile network operators (MNOs) throughout the world are now investing heavily on upgrading their existing wireless network infrastructure with fog computing. One of the key challenges for such a system is to maintain ultra-low latency and ultra-high availability between connected vehicles and fog computing networks\cite{XY2018TactileInternet}.

Due to the random nature of wireless networks, the latency performance of connected vehicles can vary significantly at different time and locations. In particular, recent results \cite{xy2019} as well as our own measurements have shown that the round-trip-time (RTT) varies substantially from second to second even at a fixed location with direct Line-of-Sight (LoS) connectivity to the base station (e.g., eNB). In other words, it is generally impossible to guarantee a deterministic latency value for a wireless connection for most of the connected vehicular systems. In addition, the fact that different vehicles may request different sets of services when driving into different locations further exacerbates the challenge. It is therefore important to establish a spatio-temporal statistic model for wireless connections between moving vehicles and a fog computing network. In this way, this model can be utilized by each connected vehicle to recover and predict the latency performances when driving to different locations. This also creates opportunities for an autonomous vehicle to warn human drivers to take over the control of vehicles when the performance of wireless connections fails to meet the requirement of the requested service at a certain location.

Spatio-temporal modeling of wireless access latency is known to be a notoriously difficult problem due to the following reasons.
First, a long-term data measuring and analysis across a wide geographical area is costy and generally difficult. Second, it is known that the number and quality of samples are
directly affecting the accuracy of the
derived model. In particular, when the number of samples is insufficient, an occasional bias can lead to a serious error. It is generally impossible to collect the sufficient number of samples across a wide temporal and spacial dimension with the same quality. Also the number of required samples to train a model with satisfactory performance can vary significantly under different scenarios and services. The gap between the insufficient number of samples that can be collected with the limited time and cost and the large amount of data required for training an accurate model needs to be filled. Finally, sythesizing a large number of samples across a wide range of time and space is notoriously difficult. Currently, there is still lacking a simple and accurate model for modeling and predicting the latency in wireless networks.

In this paper,
we introduce LaMI (Latency Model Inpainting), a novel model inpainting framework to generate a complete spatio-temporal model that can capture statistic feature of the latency performance of a connected vehicular system across a wide-geographical area. LaMI first collects some latency performance traces from vehicles driving throughout the considered areas for a certain period of time. Due to the different location popularity and population densities, the number of samples recorded in data collection may vary significantly across different locations, some of which are not sufficient to generate the empirical latency PDF with required accuracy. To address this issue, LaMI adopts the concept of inpainting from image processing in which the spatial correlation of the empirical PDF across different locations will be evaluated and samples collected in different regions with high correlation will be carefully selected and duplicated into each other. For those regions which still cannot collect sufficient numbers of samples, we propose a VAE-based approach to generate new samples according to the unknown probability distribution of the existing samples. Finally, LaMI fits the histogram of the samples collected and generated at each location and evaluate the latency and reliability performance of different services that can be provided to each connected vehicle.

Extensive simulations have been conducted based on the latency data collected in the commercial LTE network in a university campus to evaluate the performance of LaMI. Results show that LaMI can significantly improve the recovery accuracy of spatio-temporal modeling of a vehicular network especially with the latency-demanding services. In particular, compared to interpolation and nearest neighbor approaches, LaMI can significantly improve the accuracy of the empirical latency PDF. Furthermore, LaMI can be directly extended to some network models with limited or unbalanced samples across wide temporal and geographical.
\section{Related Work}
{\bfseries Connected Vehicle Networks}: Connected vehicular network, especially C-V2X, has attracted significant interest in both academia and industry \cite{5GAAUseCases, Zhang2013, Elbatt2006Cooperative, Vehicularsurvey2018, ge2015spatial}. More specifically, a multilevel information fusion approach to better process atomic messages was proposed in \cite{Zhang2013}. In \cite{Elbatt2006Cooperative}, the authors focused on vehicular safety about collision avoidance to improve the vehicular safety. The readers can refer to \cite{5GAAUseCases} for a detailed survey of the typical use cases, methodologies, and service level requirements of C-V2X.

{\bfseries Latency modeling}: 
Most existing works focused on how to ensure the latency experienced by a connected vehicle below a deterministic threshold when driving into different locations\cite{Vehicularsurvey2018}. Therefore, modeling the probability distribution function of latency is critically important. In \cite{xy2019}, the authors proposed Adaptivefog, a novel framework to maximize confidence levels in LTE-based fog computing for smart vehicles. A new scheduler was proposed in \cite{RAVEN2018} for mitigating the latency in vehicular systems. In \cite{VehicularCommunications2017}, the authors modeled some key characteristics for wireless channels in connected vehicular networks.

{\bfseries Generative modeling}: Deep generative models have already exhibited great potential in generating different kinds of complicated datasets. In \cite{VAE2014}, the authors gave a comprehensive introduction on a popular deep generative model, named VAE. In \cite{TutorialVAE2016}, the authors presented many applications based on VAE, such as generating undistinguished house numbers \cite{DBLP:GregorDGW15} and high-resolution photographic images \cite{introvae}.


 \vspace{-0.1in}
\section{Architecture Overview}
\vspace{-0.05in}
A fog computing-supported vehicular system consists of the following elements\cite{XY2018TactileInternet}:
\begin{enumerate}
  \item Vehicles: connected to the wireless communication networks to obtain different in-vehicle functions and with different latency requirements for different services request.
  \item Wireless access points (APs): belonging to a part of the network infrastructure of a mobile network operator (MNO) to provide wireless connectivity of vehicles. They will provide wireless communication links between vehicles and fog nodes. In this paper, we consider the commercial LTE connection offered by an MNO to transmit workload request and processing results between vehicles and fog nodes.
  \item Fog nodes: correspond to low-cost edge computing servers that can support local computation-intensive services for connected vehicles.
\end{enumerate}

In this paper, we focus on the spatio-temporal statistic modeling of a connected smart vehicular system in which each vehicle must always be able to evaluate and generate its latency performance to the nearest fog node while driving to different locations. The accurate evaluation and generation of the latency performance is critically important for smart vehicular systems that rely on services offered by fog computing networks to make driving decisions. In particular, a vehicle may request different sets of services when driving to different locations \cite{5GAAUseCases}.

\begin{figure}[htbp]
\centering
\includegraphics[width=8cm]{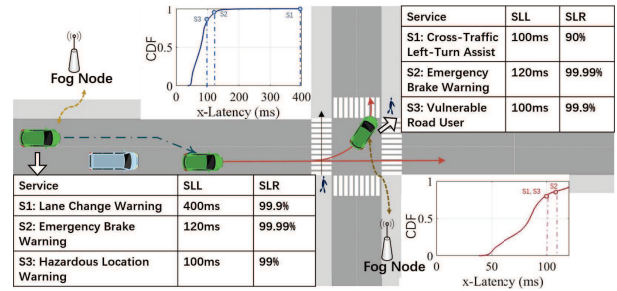}
\caption{\small{Typical connected vehicular services requested at different locations. }}
\label{fig 5GAAservice}
\end{figure}
\vspace{-0.12in}
In this paper, we assume a connected vehicle will only rely on the fog computing network to make its driving decision if the wireless connection between it and the chosen fog node can satisfy these requirements. Otherwise, the vehicle will either switch to a conservative driving mode relying on the on-board sensors or simply return the full control of the vehicle back to the human driver. In Fig. \ref{fig 5GAAservice}, we illustrate the possible services and the corresponding performance requirements as well as the empirical latency Cumulative Distribution Function (CDF) generated by our own measured dataset.

We propose a novel framework, called LaMI, to establish a complete spatial-temporal stochastic model to capture granular details of the latency performances of different services for a connected smart vehicle when driving across a wide geographical area. In particular, LaMI does not require to collect sufficient samples at every individual location and time segment. It takes advantage of the spatial correlation of the latency performance across different locations and adopts deep generative neural network to generate new samples in other under-sampled locations. LaMI consists of four main modules as shown in Fig. 2.
\begin{figure}[htbp]
\centering
\includegraphics[width=8.5cm]{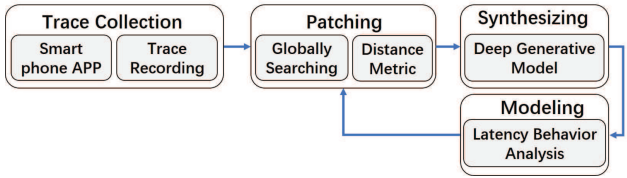}
\caption{\small{Key modules of LaMI}}
\vspace{-0.2in}
\label{fig LAMI}
\end{figure}

{\bfseries Trace collection}: Samples related to the latency performance of connected vehicles must be  collected throughout the considered area firstly. Generally speaking, the number of samples collected at different locations during different time periods can vary significantly due to the random feature of human driving behaviors and different population distributions.

{\bfseries Patching}: To estimate the latency performance at locations with insufficient numbers of samples or without samples, LaMI searches for regions that have similar latency performance and the samples can be shared across these regions.

{\bfseries Synthesizing}: For the locations still cannot collect enough samples, we adopt a deep generative neural network to learn an approximate probability distribution from existing samples and generate new samples from the learned distribution.

{\bfseries Modeling}: An empirical spatial-temporal stochastic model will be established according to the collected as well as generated samples. The established model can be used to estimate the confidence levels of different supported services when driving to each specific location as well as further improve the prediction accuracy of the latency performance as shown in Fig. \ref{fig LAMI}.
\vspace{-0.15in}
\section{Methodology}
\subsection{Trace Collection}
\subsubsection{Data Description}
We evaluate the service latency by measuring the round-trip time (RTT) between a connected vehicle and its fog server. We have developed a trace recording APP called {\it Delay Explorer} using Android API, to periodically ping the IP addresses of a cloud server and a fog node and record their RTTs at every 500ms. We follow a commonly adapted setting and assume that the fog server is located in service gateway (S-GW). According to \cite{ePC2017}, S-GW is a more ideal location for installing computing server compared to eNodeB (eNB) or Remote Radio Unit (RRU). In addition to the RTT, {\it Delay Explorer} also records time stamps, GPS coordinates, driving speed, network connection types, etc. This is because fog server usually have high power consumptions and stable power supply, which are typically unavailable at eNBs and RRUs.
We used several smart phones installed with {\it Delay Explorer} that accumulatively collected more than 760,000 samples
in a university campus in Wuhan, China.
The trace consists of recordings from both fixed locations and driving on roads. The detailed information about the trace is given in Table \ref{tab HUST1}.

%
\begin{table} [h]
\centering
\footnotesize
\caption{Summary of the collected dataset} 
\setlength{\tabcolsep}{1mm}{
\begin{tabular} {|c|c|} 
\hline
Location & a University Campus\\ 
\hline 
Number of samples & 760,000\\
\hline
Duration of measurement & Mar 16, 2019-Apr 7, 2019\\
\hline
Size of the campus& 1153 acres\\
\hline

\end{tabular}}
\label{tab HUST1}
\end{table}

\begin{table}[h]
\footnotesize
\centering
\caption{Statistical characteristics of the collected trace}
\setlength{\tabcolsep}{2mm}{
\begin{tabular}{ccccc}
\hline
Status & Empirical PDF & Mean(ms)& STD(ms)& Median(ms)\\
\hline
Fixed & \begin{minipage}{0.12\textwidth}

      \includegraphics[width=20mm, height=12mm]{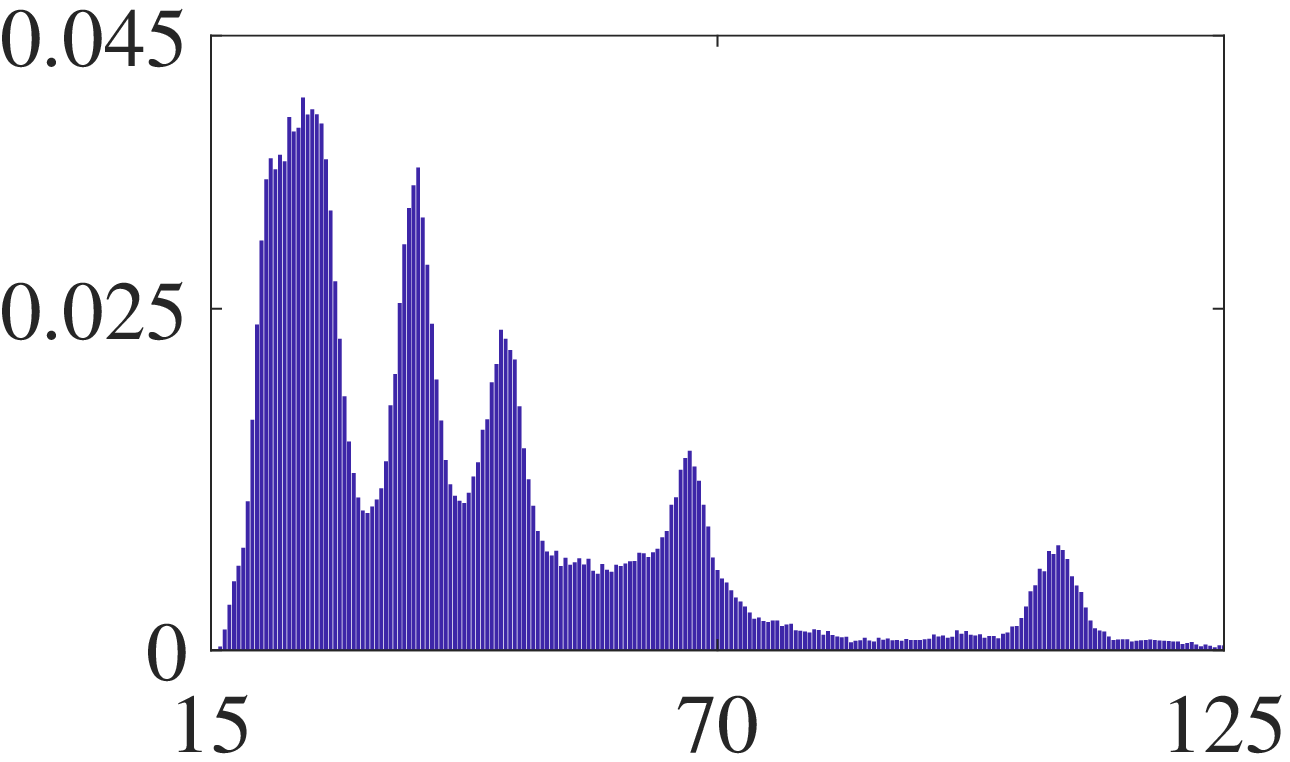}

    \end{minipage} & 43.9350& 24.7486& 37.2690\\
\hline
Moving &  \begin{minipage}{0.12\textwidth}

      \includegraphics[width=20mm, height=12mm]{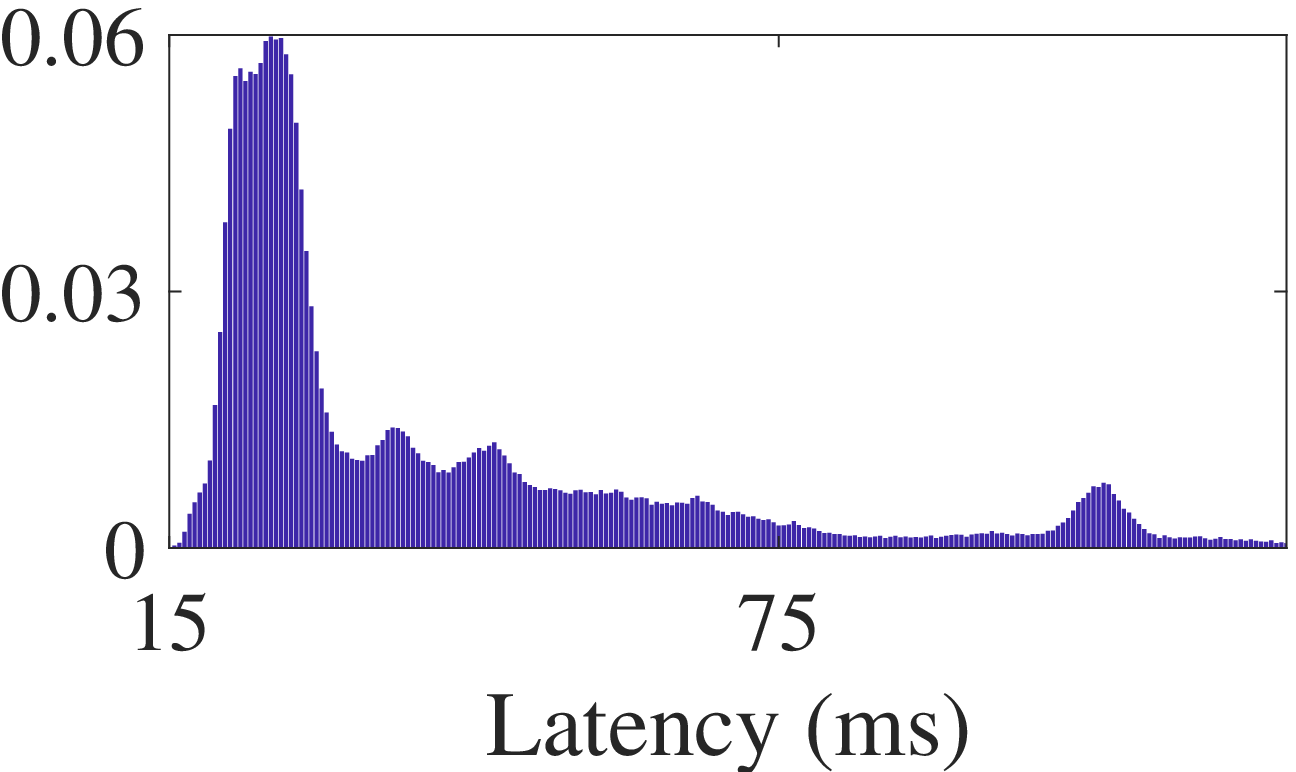}

    \end{minipage} & 43.6104 & 28.9172& 29.1490\\
\hline
\end{tabular}}
\vspace{-0.2in}
\label{tab HUST2}
\end{table}

\subsubsection{Preliminary Results and Observations}
\begin{figure}[htbp]
\centering
\includegraphics[width=5cm]{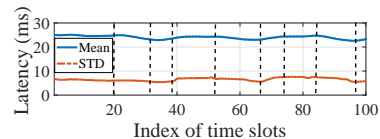}
\caption{\small{A three-day continuous RTT measurements at a fixed Lab location inside of a university campus.}}
\vspace{-0.2in}
\label{fig mean-std}
\end{figure}

Typically, the latency performance of a connected smart vehicle can be affected by its driving speed, base station location, geographical environment (e.g. blockage). As a result, it varies significantly at different locations and with time. We present the histogram, mean, standard deviation (STD), and median values of the collected trace from both fixed locations and driving recordings in Table \ref{tab HUST2}. We can observe that the mobility of a vehicle has a very limited influence on the RTT in terms of average RTT, but can lead to a much bigger increase in terms of STD.
The collected traces also imply that the temporal and spatial correlation between specific RTT values are generally very small. Even at the same location, there is no observable correlation between two consecutive samples with a 0.5s interval recorded by the same vehicle. However, the PDF of RTTs recorded at each specific location in a certain time duration can be treated as stationary as shown in Fig. \ref{fig mean-std}.

In addition, the PDF of RTTs always follows a dual-modal mixed Gaussian distribution. We generate an empirical PDF at a fixed location with different number of RTT recordings at Wuhan, China by creating histograms for the samples and fitting them with Gaussian kernel functions as shown in Fig. 4. We can observe that the empirical PDF of RTTs follows a Gaussian mixture model with two major components centered at around 25ms and 105ms. This observation aligns with recent studies \cite{RAVEN2018,ePC2017} as well as our previous work \cite{xy2019}. It implies that the dual model may be caused by the scheduling request (SR) channel periodicity and hybrid automatic repeat request (ARQ) retransmission delay. To evaluate the impact of under-sampling on the accuracy of predicted latency performance, we use the weighted KR distance \cite{xy2019} as the main metric to quantify the difference between two empirical PDFs generated with different numbers of samples. The weighted KR distance between two empirical PDFs whose CDFs are $F_i(t)$ and $F_j(t)$ is defined as:
\begin{eqnarray}
K(F_i,F_j) = \int_0^\infty\omega(t)\Delta_{ij}(t)dt,
\label{eq KRdis}
\end{eqnarray}
where $\Delta_{ij}(t)=|{F_i(t)-F_j(t)}|$ is the absolute value of the difference between two CDF values at $t$. $\omega(t)$ can adjust the importance of different latency values when comparing different CDFs, which makes KR distance more flexible in dealing with diverse service requirements.

\begin{figure}[htbp]
\centering
\subfigure[]{
  \begin{minipage}[t]{0.3\linewidth}
   \centering
   \includegraphics[width=2.8cm]{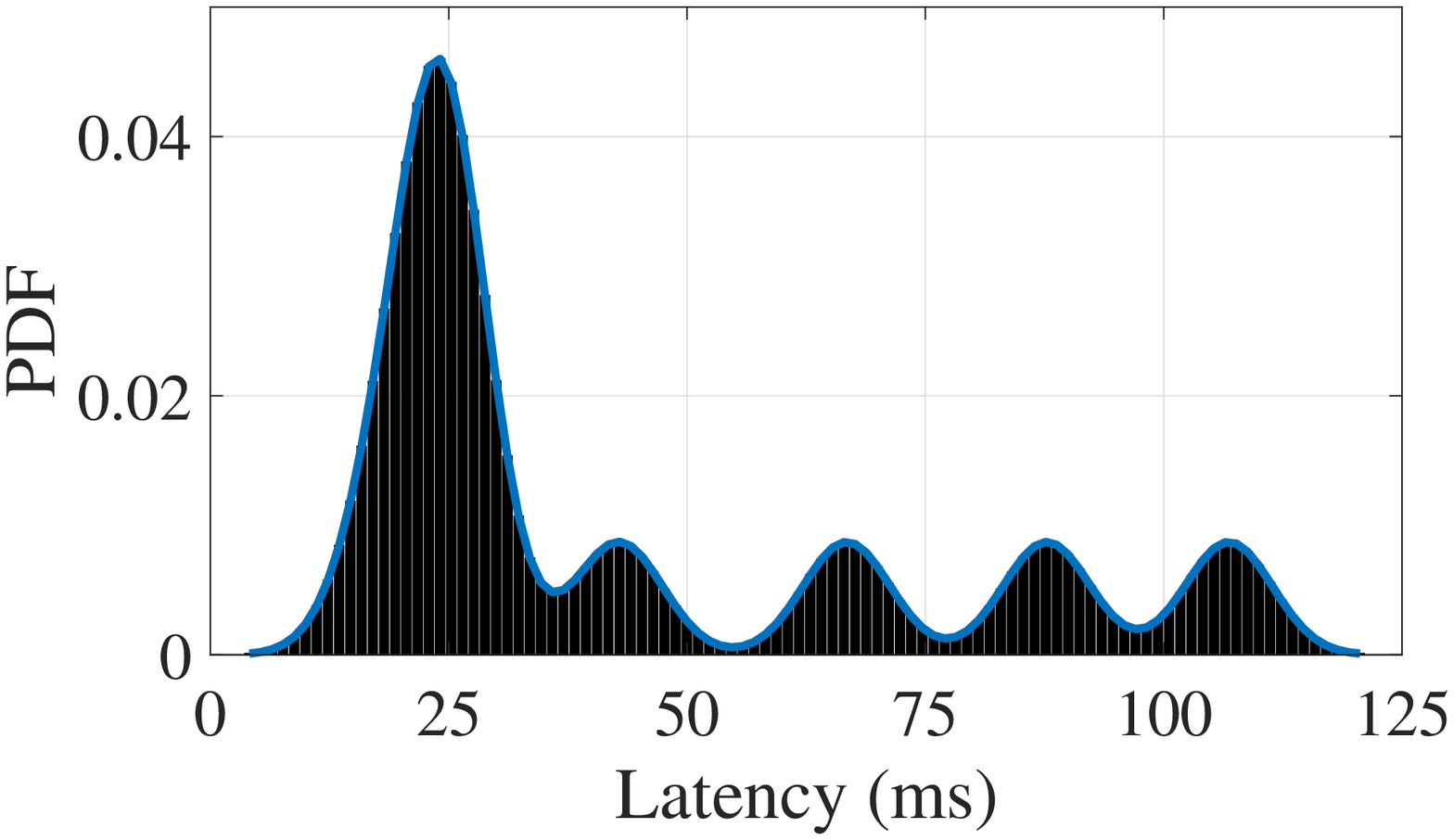}
  \end{minipage}%
}
\subfigure[]{
  \begin{minipage}[t]{0.3\linewidth}
   \centering
   \includegraphics[width=2.8cm]{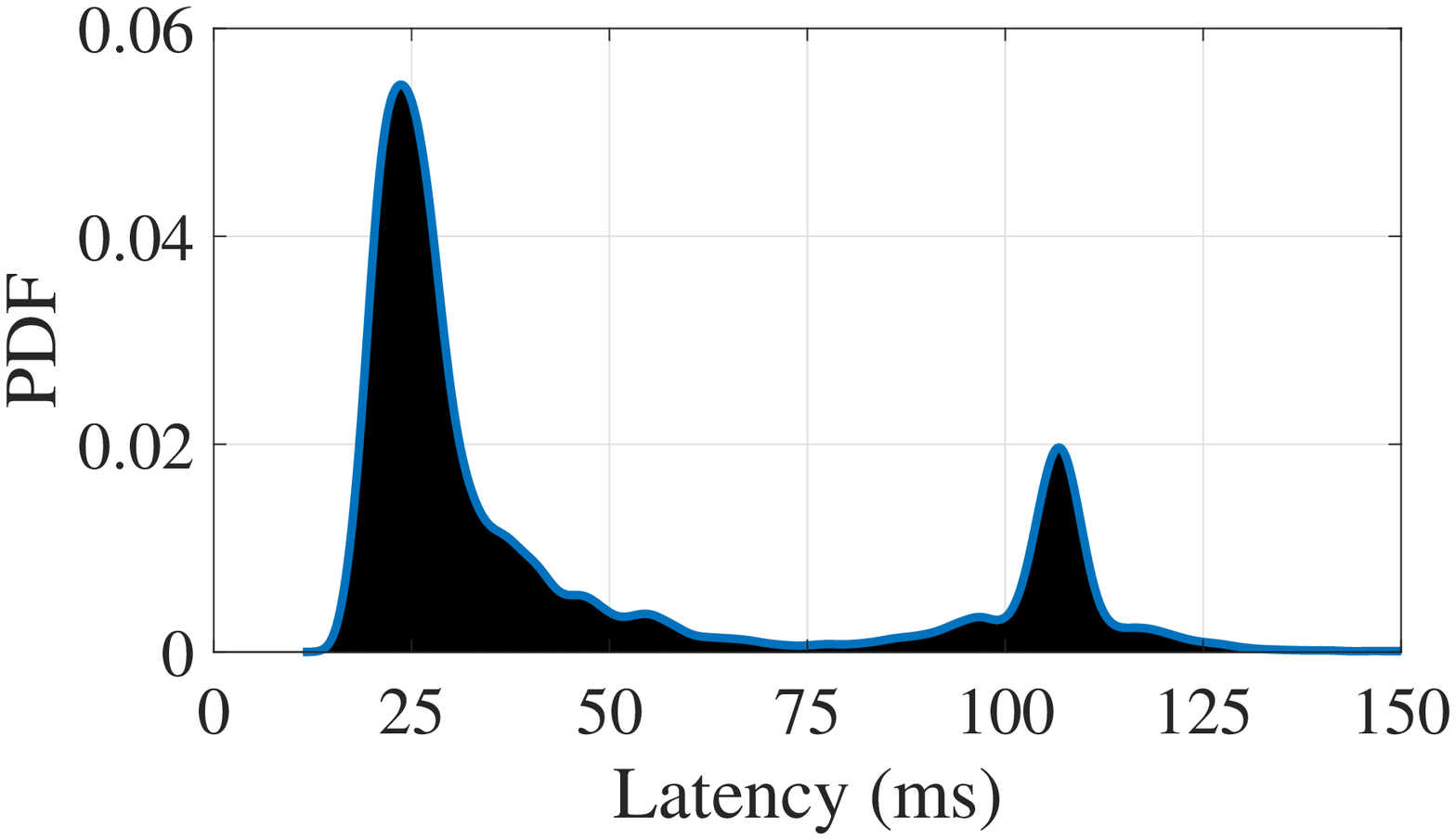}
  \end{minipage}%
}
\subfigure[]{
  \begin{minipage}[t]{0.3\linewidth}
   \centering
   \includegraphics[width=2.8cm]{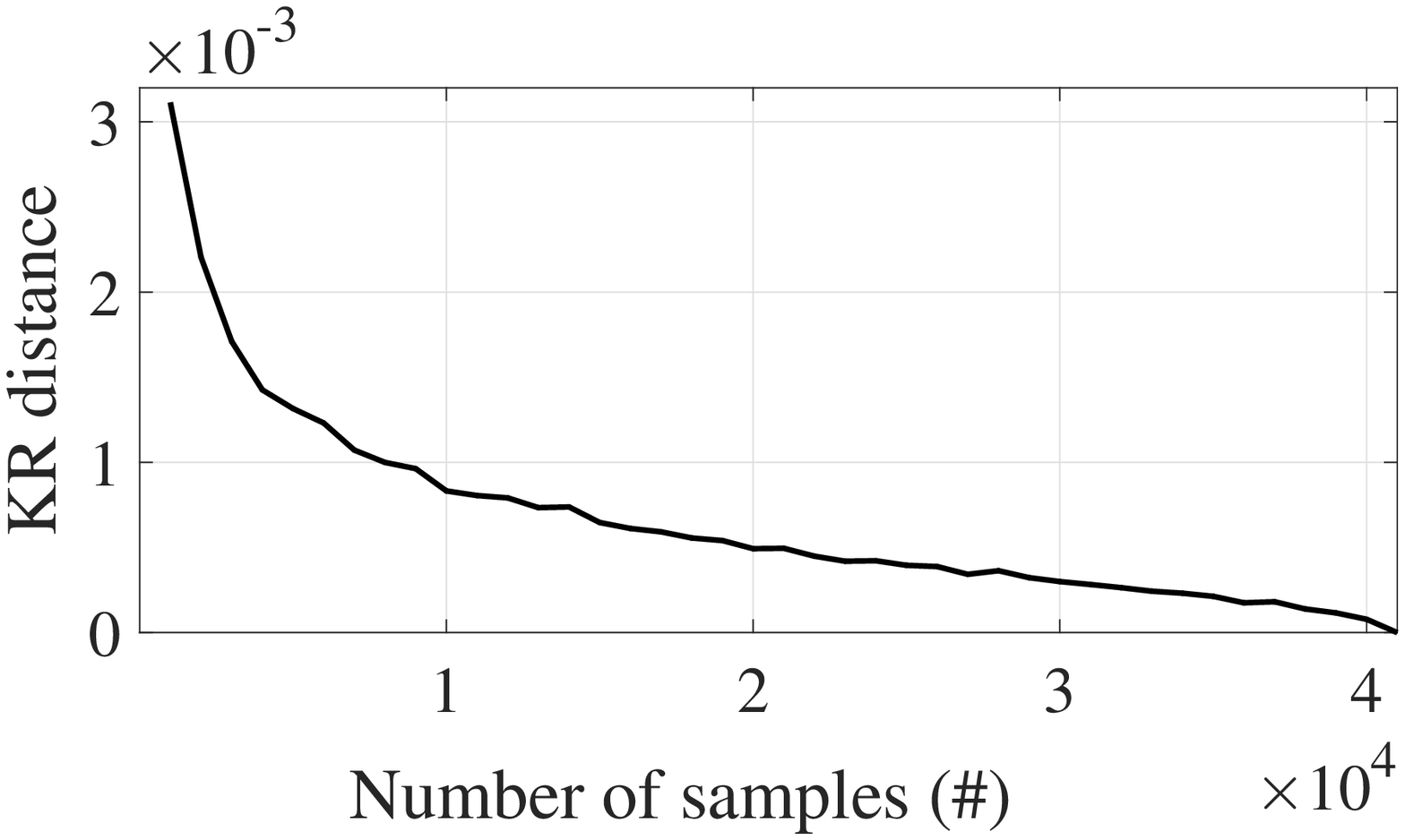}
  \end{minipage}%
}
\vspace{-0.12in}
\caption{\small{(a) Empirical PDF generated by Gaussian kernel with 10 samples, (b) Empirical PDF generated by Gaussian kernel with 40,000 samples, (c)  KR distance between empirical PDFs generated with different numbers of samples.}}
\label{fig multi-double-peak}
\end{figure}

In Fig. \ref{fig multi-double-peak}(a), we present the Gaussian kernel function fitting result of 10 samples randomly selected from the total 40,000 samples. In Fig. \ref{fig multi-double-peak}(b), we present the same fitting of the total 40,000 samples. In Fig. \ref{fig multi-double-peak}(c), we further compared the KR distance between empirical PDF generated with different numbers of samples and that generated with all 40,000 samples collected in a fixed location. We can observe that with a larger number of samples we can generate an empirical PDF with higher accuracy. In particular, with the increase of the number of samples, KR distance will first decrease rapidly, and then slow down.

\begin{figure}[htbp]
\centering
\includegraphics[width=4.5cm]{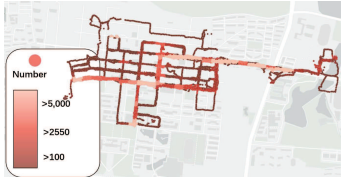}
\caption{\small{The sample locations and volumes across a university campus.}}
\vspace{-0.3in}
\label{fig shadeHUST}
\end{figure}

We can also observe that the number of samples collected at different locations exhibits strong unbalance, with smaller number of samples collected in remote areas and much higher density of samples recorded in urban areas. In Fig. \ref{fig shadeHUST}, we use different shades to show the number of samples collected at each location for all four traces.
\vspace{-0.12in}
\vspace{-0.12in}
\vspace{-0.0212in}
\subsection{Patching}
As mentioned before, in some sparsely populated areas, it is difficult to record a sufficient amount of samples to establish an accurate empirical PDF for latency performance predicting. Most existing solutions focus on estimating the performance of the under-sampled region directly using samples or performances from the nearest neighbors. In this paper, we adopt the concept of inpainting from image processing and propose a global searching and patching approach. In this approach, we will first search for areas whose PDFs have the smallest weighted KR distance with the under-sampled region, and then patch the under-sampled region with the samples from the chosen areas.

Our approach is based on semantic image inpainting and synthesis methods that have been successfully used in fixing and reconstructing images with missing regions, which is also known as exemplar-based image inpainting. In this approach, if one or more patches are identified as similar to the missing or damaged part of the image, chunks of these patches will be copied and pasted to reconstruct the missing region.

Unfortunately, exemplar-based image inpainting approaches cannot be directly applied to complete the spatio-temporal model of latency performance due to the following reasons:
\begin{enumerate}
  \item Due to the inaccuracy of GPS locations, it is generally impossible to identify the nearest neighbors of the under-sampled locations. In other words, the $k$-nearest neighboring-based approaches cannot be directly applied to reconstruct the latency traces for the under-sampled area.
  \item Instead of recovering individual pixel values for the missing regions in image processing, our problem requires to reconstruct the PDF for each location. As mentioned before, RTT is a stochastic value that varies sharply at different time and locations. We focus on reconstructing the empirical PDF of the RTT for a connected smart vehicle. In other words, Euclidean distance-based metrics cannot be used to evaluate the similarity of different regions (e.g. patches).
  \item Driving latency can be affected by many external factors such as driving speed, weather, and locations of network infrastructure, all of which should be considered when comparing the similarity between different patches.
\end{enumerate}

To address the above issues, we propose a novel patching approach based on statistic distance between various regions. In particular, to address issue 1), we divide the entire area of sample collection into a number $N$ of equal-sized regions. Vehicles will experience the same PDF when driving within each region. 
To address issue 2), we measure the similarity of different regions using statistic distance metric between their PDFs. In particular, suppose $\Phi_p$ and $\Phi_q$ are the sets of samples in regions $p$ and $q$, respectively. In this paper, we consider the weighted KR distance metric with the weight factor $\omega(t)$ = 1 to evaluate the similarity between two regions $\Phi_p$ and $\Phi_q$.
%
To address issue 3), we introduce a weighting factor, denoted as $\alpha_{pq}$, to capture the similarity of environmental elements such as surrounding buildings, types of roads (highway or any incity-road) as well as other features between two regions. Suppose region $p$ is the region without sufficient number of samples and can be patched with the samples from region $q$. In this case, the value of $\alpha_{pq}$ will be used to calculate the size of the samples that will be duplicated from region $q$ to patch region $p$. 

\begin{algorithm}[h]
\setstretch{1.3}
 \footnotesize
 \caption{Patching}\label{Algorithm 1}
  \begin{itemize}
   \item[]  {\bf Input}: Objective region $p$
  \begin{itemize}
   {\bf if} Number of samples in $p\leq M$:
   \begin{itemize}
    \item[] Search for candidate patches:
    \item[] \quad $-$ Find the region $\hat q$: $\hat q$=$\mathop{\arg\min}\limits_{q\in {\Omega \backslash \Phi_p}}$
        \begin{eqnarray}
        K\left( CDF(\Phi_p\bigcup \Psi_p), CDF(\Phi_q\bigcup \Psi_q)\right) \nonumber
        \end{eqnarray}
        \quad where $CDF(\cdot)$ is the empirical CDF generated with the

        \quad given samples;
    \item[] \quad $-$ Duplicate a portion of $\Phi_{\hat q}$ to $\Phi_{\hat p}$. The portion size is evaluated according to $\alpha_{p\hat q}$;
    \item[] \quad $-$ Update $\Omega$=$\Omega$-$\hat q$;
   \end{itemize}
   {\bf else}:
   \item[] \quad Do not search, return $\Phi_p$;
  \end{itemize}
  \item[]  {\bf Output}: Updated $\Phi_{\hat p}$;
 \end{itemize}
\end{algorithm}
In Algorithm \ref{Algorithm 1}, we try to recover and improve the empirical PDFs for regions with less than $M$ number of samples. Let us define $\Omega = \bigcup_{r=\{1, \ldots, N\}} \Phi_{r}$ as the set of all the samples 
collected in $N$ regions. Suppose we try to recover the empirical PDF of region $p$. Let $\Phi_p$ be the set of samples in region $p$. $\Psi_p$  is the combination of all the samples in the nearest neighboring regions of region $p$. Then a global searching for candidate patches, regions with similar PDF with region $p$, will be performed by evaluating the KR distance between combination of $\Phi_p$ and $\Psi_p$ and all the other similar combinations of regions throughout the entire area. 
We use $\Phi_{\hat p}$ to denote the updated set of samples including the original samples and the duplicated samples from the candidate patches. 

\vspace{-0.092in}
\subsection{Synthesizing}
There are locations that still cannot have sufficient numbers of samples to generate an accurate empirical PDF. In this case, we adopt VAE, a deep generative neural network approach to generate more new samples. Compared to other neural network-based approaches such as regression, VAEs are generative models that can output new random samples that has similar distribution with the input data, i.e., the samples collected or patched at a given location. In this paper, we focus on generating new samples that have similar distribution with the original and patched samples at location $p$. To achieve this, VAE introduces a latent variable that can influence the model to generate new artificial samples that are similar to the existing samples in $\Phi_{\hat p}$. We use $z$ to denote the vector of latent variables  that are sampled according to PDF $P(z)$. Our main objective is to find the PDF of samples in location $p$, defined as $P(x)$ for $x\in \Phi_{\hat p}$. Following the law of probability, we can write the relationship between $x$ and $z$ as follows:
\begin{eqnarray}
\begin{split}
P(x)=\int P(x|z)P(z)dz.
\end{split}
\label{eq_Px}
\end{eqnarray}

To infer the distribution $P(x|z)$, we adopt the classical variational inference approach to approximate the real distribution $P(z|x)$ using some typical distributions, e.g., Gaussian distribution. In particular, we introduce a Gaussian distribution function $Q(z|x)$ and try to infer $P(z|x)$ from $Q(z|x)$ by minimizing the KL divergence between these two distributions. According to its definition, the KL divergency between $P(z|x)$ and $Q(z|x)$ is given by
\begin{eqnarray}
D_{KL}[Q(z|x)||P(z|x)]\overset{A}{=} & E_{z\thicksim Q}[\log Q(z|x)-\log P(z|x)],\\
\overset{B}{=} & E_{z\thicksim Q}[\log Q(z|x)-\log P(x|z) \nonumber\\
&-\log P(z)]+\log P(x),
\label{eq_DKLQP}
\end{eqnarray}
where the equation sign (A) is according to the definition of KL divergence and (B) is established from the fact that the expectation over $z$ will affect $P(x)$ and we can therefore move $\log P(x)$ outside of the expectation. The right-hand-side of (\ref{eq_DKLQP}) consists of another KL divergency, and we have
\begin{eqnarray}
\lefteqn{\log P(x)-D_{KL}[Q(z|x)||P(z|x)]} \nonumber\\
&&=E_{z\thicksim Q}[\log P(x|z)] -D_{KL}[Q(z|x)||P(z)].
\label{eq final}
\end{eqnarray}

Note that (\ref{eq final}) consists of two distribution functions $Q(z|x)$ to project samples $x$ into a latent variable space and $P(x|z)$ to generate new samples with given latent variable $z$. We can then adopt the autoencoder to train a neural network with $Q(z|x)$ as the encoder net and $P(x|z)$ as the decoder net.

Another way to interpret equation (\ref{eq final}) is that VAEs will try to find $\log P(x)$ under error $D_{KL}[Q(z|x)||P(z|x)]$. In other words, VAEs will try to minimize the lower bound of $\log P(x)$, i.e., we have
\begin{eqnarray}
\begin{split}
\log P(x)\geqslant \mathbb{L}(x)&=E_{z\thicksim Q}[\log P(x|z)]\\
&-D_{KL}[Q(z|x)||P(z)].
\end{split}
\label{eq lowerbound}
\end{eqnarray}

As shown in \cite{VAE2014}, this minimized lower-bound is a good approximation of $\log P(x)$ especially considering that the exact distribution of $P(x)$ is generally untractable.

\vspace{-0.17in}
\subsection{Modeling}

After creating a sufficient number of samples, an empirical PDF can be established by fitting the histogram of the RTTs using the Gaussian kernel function. The empirical PDF can be used to estimate the confidence levels for different services.

Our generated PDF is an approximation of the PDF of the real service latency which is critical to the safety-related services requiring stringent latency performance guarantees. In this case, we can introduce a weighting factor $\omega_r$ to different segment of the generated empirical CDF according to the required latency values of safety related services. Suppose the maximum tolerable latency of vulnerable road user service is given by 100ms, we can write the adjusted confidence of supporting vulnerable road user service with the required latency level as
\begin{eqnarray}
f(x\leq r)=\omega_r P(x\leq r),
\label{eq modeling}
\end{eqnarray}
where $r=100$ms and $0\leq\omega_r\leq1$.

\vspace{-0.1in}
\section{Numerical Results}

\begin{figure}[htbp]
\centering
\subfigure[]{
  \begin{minipage}[t]{0.43\linewidth}
   \centering
   \includegraphics[width=4.cm]{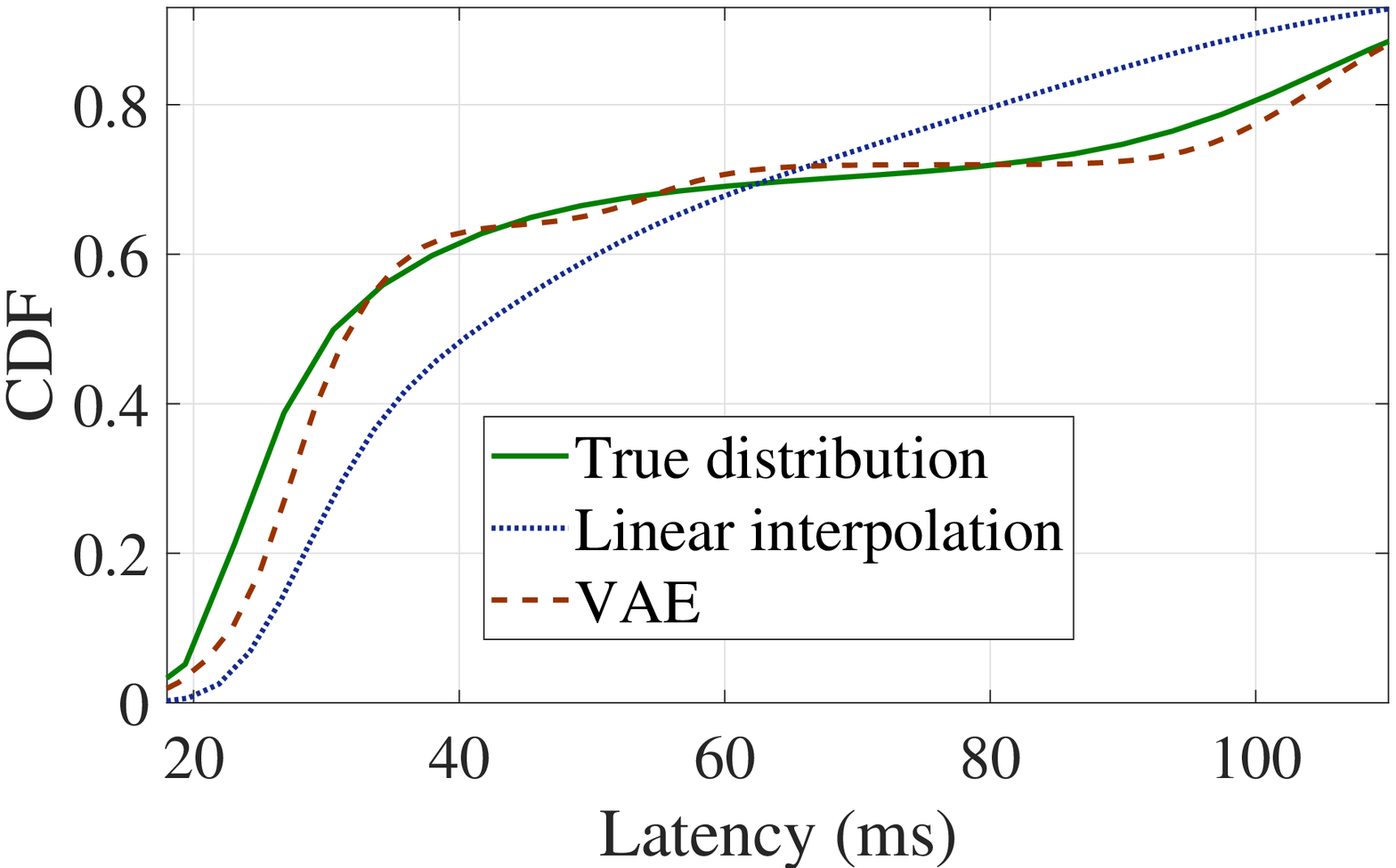}
  \end{minipage}%
}
\subfigure[]{
  \begin{minipage}[t]{0.43\linewidth}
   \centering
   \includegraphics[width=4.cm]{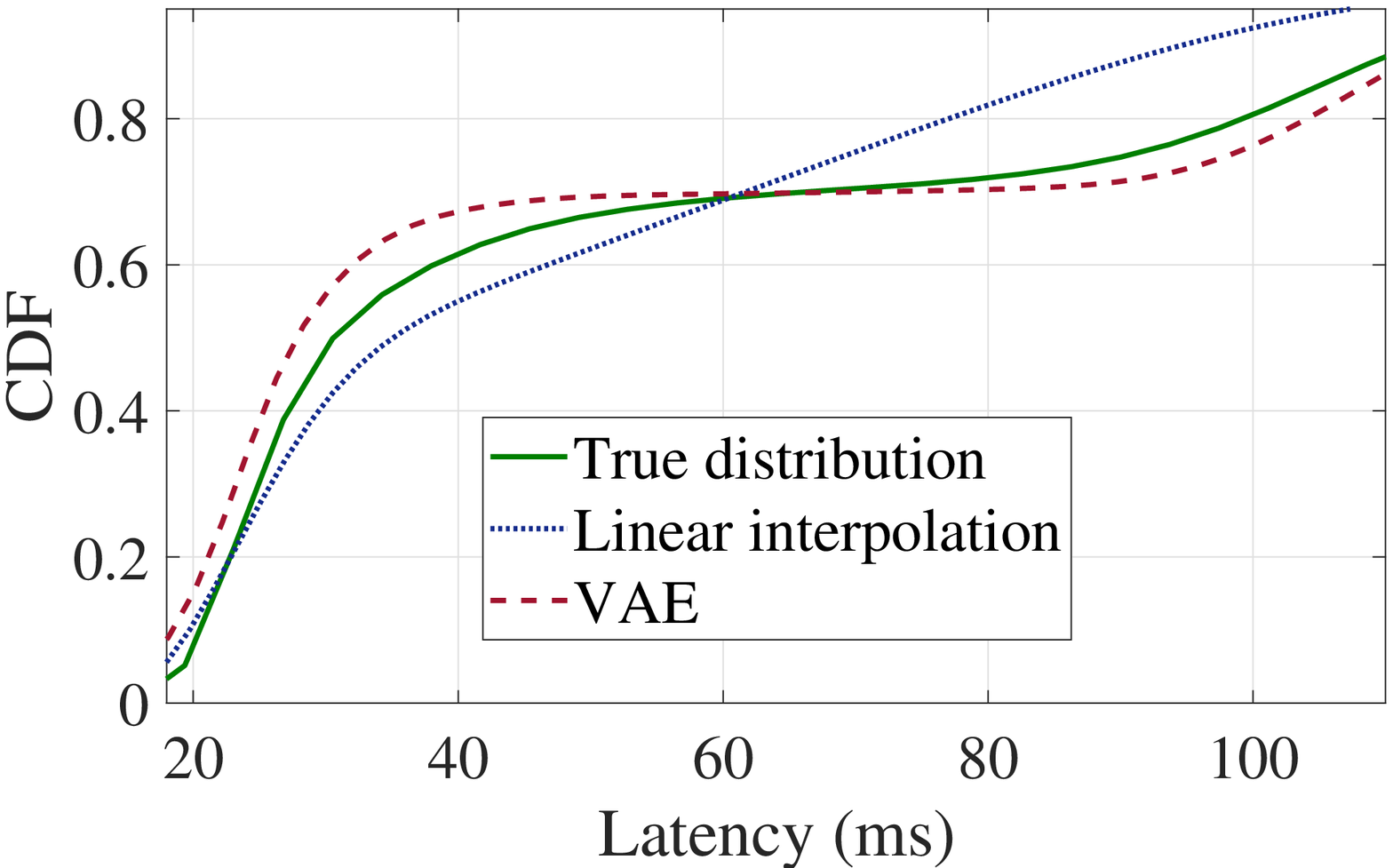}
  \end{minipage}%
}
\vspace{-0.1in}
\caption{\small{(a) Empirical CDF recovered by VAE with 25 samples and (b) Empirical CDF recovered by VAE with 10 samples in the same location, compared with true distribution generated by 50,000 samples.}}
\vspace{-0.1in}
\label{simu VAE}
\end{figure}

In this section, we evaluate the performance of LaMI using dataset collected in a university campus. The sample collecting locations and volumes are shown in 
Figure \ref{fig shadeHUST}.

As mentioned earlier, the total number of samples directly affected the accuracy of the recovered models especially when the number of samples is small. We therefore compare the sample recovery performance of VAE and linear interpolation, one of the popular sample recovery approaches in Figure \ref{simu VAE}.
In particular, we present the empirical CDF recovered by VAE with a very small number of samples (e.g., 10 and 25 samples) with the true distribution generated by all 50,000 samples collected in the same location.
We can observe that the result obtained by interpolation tends to weaken the peak of PDF, which may cause a larger bias between the two peaks at around 90ms. In contrast, VAE outperforms the interpolation approach and can better capture the properties of the true latency distribution. 

\begin{figure}[htbp]
\centering
\subfigure[]{
  \begin{minipage}[t]{0.45\linewidth}
   \centering
   \includegraphics[width=4cm]{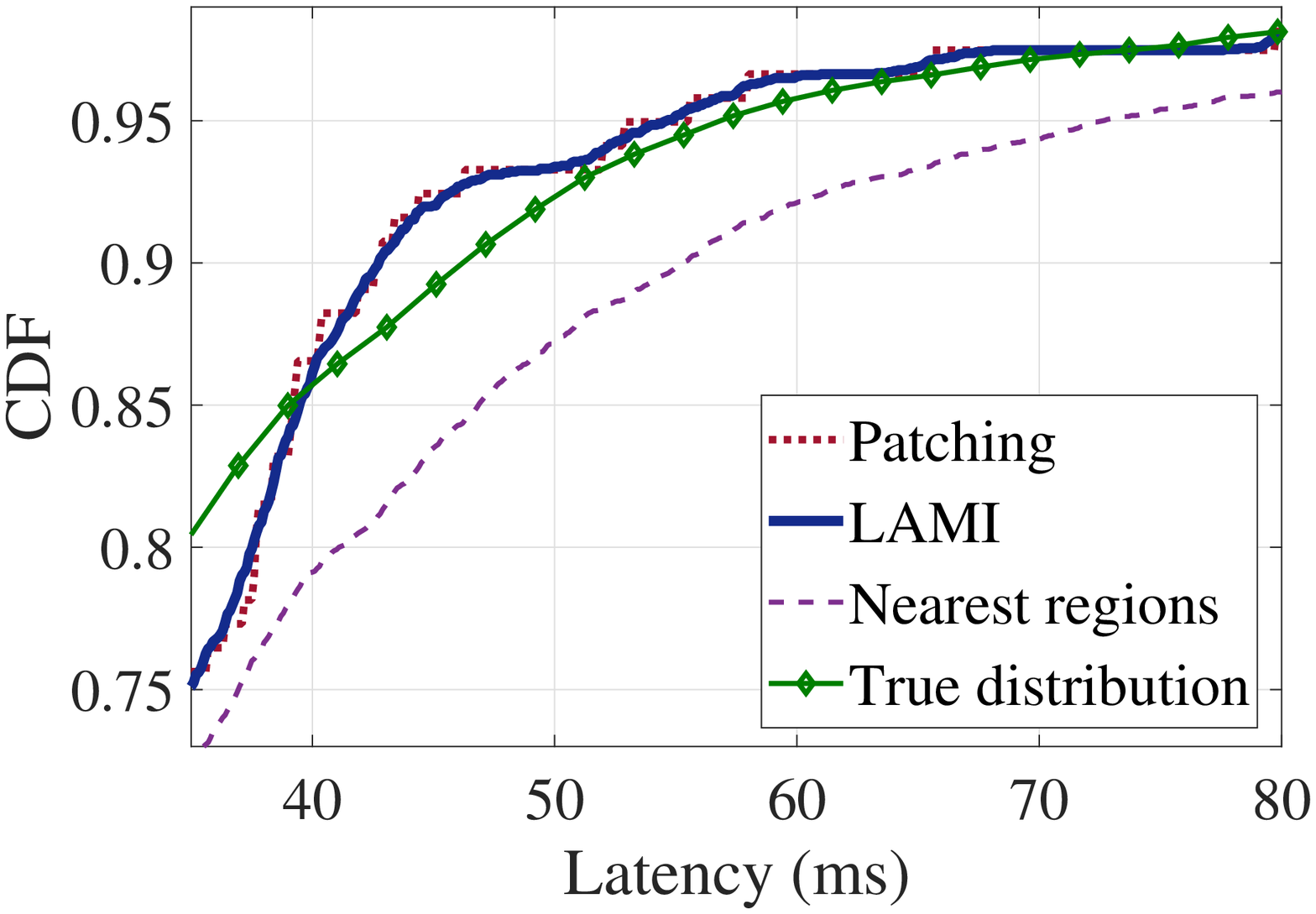}
  \end{minipage}%
}
\subfigure[]{
  \begin{minipage}[t]{0.45\linewidth}
   \centering
   \includegraphics[width=4cm]{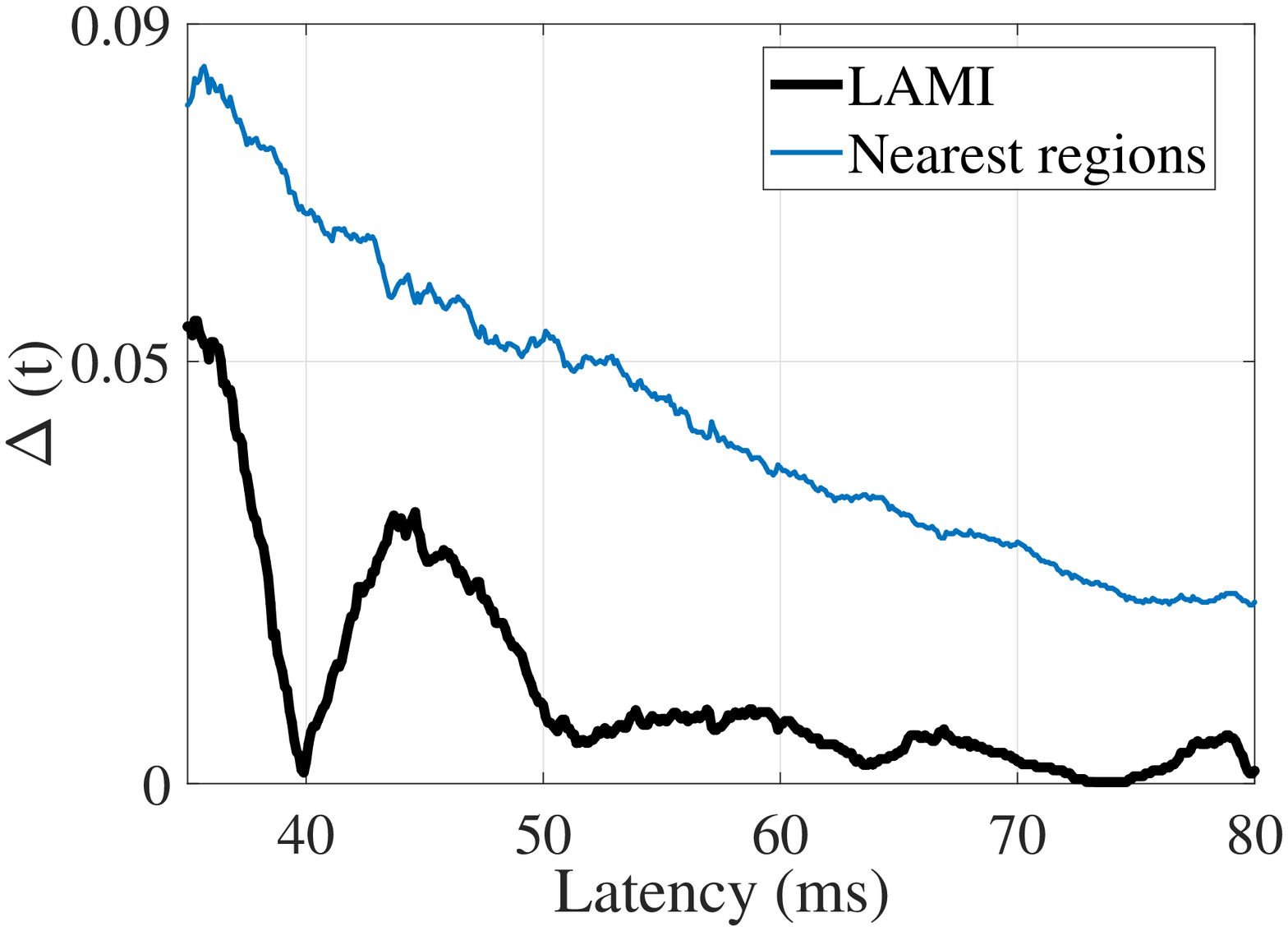}
  \end{minipage}%
}
\vspace{-0.15in}
\caption{\small{(a) Empirical CDFs recovered by various methods compared with the true distribution and (b) Absolute difference value between the recovered empirical CDFs and the true distribution CDF.}}
\label{simu LAMI}
\end{figure}
\vspace{-0.1in}
To evaluate the sample recovery performance of two main models of LaMI: patching and synthesizing, we consider a special scenario and assume no samples have been recorded for a considered area. In this case, LaMI needs to first perform the global search for candidate patches, and then duplicate some high correlated samples collected in other areas. Finally, LaMI applies VAEs with the patched samples to generate empirical CDFs. Experimental results are presented in Fig. \ref{simu LAMI}. We can observe that LaMI achieves a much better performance compared to the nearest neighboring approach, a commonly used approach that generates empirical CDF with samples from the neighboring regions. 

\begin{figure}
\centering
\includegraphics[width=4cm]{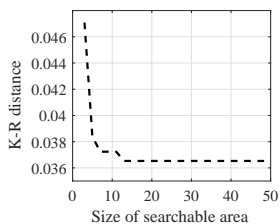}
\vspace{-0.1in}
\caption{\small{KR distances under different sizes of searchable area.}}
\vspace{-0.1in}
\label{simu Omega}
\end{figure}

In Figure \ref{simu Omega}, we evaluate the impact of the searchable area size on the performance of LaMI. In particular, we compare the KR distance between the recovered empirical PDF and true distribution under different sizes of total searchable areas. 
We can observe that with the increasing of the searchable area for candidate patches, the KR distance between the empirical PDF recovered by LaMI and the true distribution reduces.   
The KR distance however will approach to a constant when the searchable area continues to expand. This means that a larger searchable area does not always result in better performance.  
Note that with the searchable area continuing to increase, the misjudgment rate can also raise. In other words, the searchable area should be carefully decided to balance the improvement on model accuracy and the recovery error.


\vspace{-0.1in}
\section{Conclusion}
In this paper, we have proposed a novel approach to model the spatio-temporal latency performance for connected vehicular networks. In particular, we have collected RTT measurements between vehicles and fog nodes through commercial LTE networks in a university campus. Based on the thorough analysis of the collected dataset, a novel spatio-temporal generative model named as LaMI has been proposed to handle the challenges in the latency modeling in smart vehicular systems. LaMI can inpaint the missing latency samples by searching for similar regions and sharing their samples. A deep generative model based on VAE is also adopted to further improve the modeling accuracy. Numerical results show that the proposed LaMI framework can significantly improve the modeling accuracy of latency performance compared to existing popular solutions such as linear interpolation and nearest neighbor-based method. 

\vspace{-0.1in}
\section*{Acknowledgment}
The authors would like to thank Ericsson (China) Hubei Branch and  China Mobile Hubei 5G Joint-innovation Lab for help in the data collection. 

\vspace{-0.15in}
\bibliographystyle{IEEEtran}
\bibliography{bib21}

\begin{thebibliography}{10}
\providecommand{\url}[1]{#1}
\csname url@samestyle\endcsname
\providecommand{\newblock}{\relax}
\providecommand{\bibinfo}[2]{#2}
\providecommand{\BIBentrySTDinterwordspacing}{\spaceskip=0pt\relax}
\providecommand{\BIBentryALTinterwordstretchfactor}{4}
\providecommand{\BIBentryALTinterwordspacing}{\spaceskip=\fontdimen2\font plus
\BIBentryALTinterwordstretchfactor\fontdimen3\font minus
  \fontdimen4\font\relax}
\providecommand{\BIBforeignlanguage}[2]{{%
\expandafter\ifx\csname l@#1\endcsname\relax
\typeout{** WARNING: IEEEtran.bst: No hyphenation pattern has been}%
\typeout{** loaded for the language `#1'. Using the pattern for}%
\typeout{** the default language instead.}%
\else
\language=\csname l@#1\endcsname
\fi
#2}}
\providecommand{\BIBdecl}{\relax}
\BIBdecl

\bibitem{Market}
\BIBentryALTinterwordspacing
{Market Research Engine}, ``Global connected cars market - trends and forecast:
  2015-2020.'' [Online]. Available:
  \url{https://www.marketresearchengine.com/reportdetails/global-connected-cars-market}
\BIBentrySTDinterwordspacing

\bibitem{XY2018TactileInternet}
Y.~Xiao and M.~Krunz, ``Distributed optimization for energy-efficient fog
  computing in the tactile {I}nternet,'' \emph{IEEE J. Sel. Area Commun.},
  vol.~36, no.~11, pp. 2390--2400, Nov. 2018.

\bibitem{xy2019}
Y.~{Xiao}, M.~{Krunz}, H.~{Volos}, and T.~{Bando}, ``Driving in the fog:
  Latency measurement, modeling, and optimization of lte-based fog computing
  for smart vehicles,'' in \emph{IEEE SECON}, Boston, MA, Jun. 2019.

\bibitem{5GAAUseCases}
G.~WhitePaper, ``5gaa c-v2x use cases methodology, examples and service level
  requirements,'' June. 2019.

\bibitem{Zhang2013}
L.~Zhang, D.~Gao, W.~Zhao, and H.~C. Chao, ``A multilevel information fusion
  approach for road congestion detection in vanets,'' \emph{Mathematical and
  Computer Modelling}, vol.~58, no. 5-6, pp. 1206--1221, Sep. 2013.

\bibitem{Elbatt2006Cooperative}
T.~ElBatt, S.~K. Goel, G.~Holland, H.~Krishnan, and J.~Parikh, ``Cooperative
  collision warning using dedicated short range wireless communications,'' in
  \emph{Proceedings of the 3rd international workshop on Vehicular ad hoc
  networks}, Los Angeles, CA, Sep. 2006.

\bibitem{Vehicularsurvey2018}
S.~K. Bhoi and P.~M. Khilar, ``Vehicular communication: a survey,'' \emph{IET
  networks}, vol.~3, no.~3, pp. 204--217, Sep. 2014.

\bibitem{ge2015spatial}
X.~Ge, B.~Yang, J.~Ye, G.~Mao, C.-X. Wang, and T.~Han, ``Spatial spectrum and
  energy efficiency of random cellular networks,'' \emph{IEEE Tran. Commun.},
  vol.~63, no.~3, pp. 1019--1030, Mar. 2015.

\bibitem{RAVEN2018}
H.~Lee, J.~Flinn, and B.~Tonshal, ``{RAVEN}: Improving interactive latency for
  the connected car,'' in \emph{ACM MobiCom}, New Delhi, India, Oct. 2018.

\bibitem{VehicularCommunications2017}
L.~{Liang}, H.~{Peng}, G.~Y. {Li}, and X.~{Shen}, ``Vehicular communications: A
  physical layer perspective,'' \emph{IEEE Transactions on Vehicular
  Technology}, vol.~66, no.~12, pp. 10\,647--10\,659, Dec. 2017.

\bibitem{VAE2014}
\BIBentryALTinterwordspacing
D.~P. Kingma and M.~Welling, ``Auto-encoding variational bayes,'' \emph{arXiv},
  Dec. 2013. [Online]. Available: \url{https://arxiv.org/abs/1312.6114}
\BIBentrySTDinterwordspacing

\bibitem{TutorialVAE2016}
C.~Doersch, ``Tutorial on variational autoencoders,'' \emph{arXiv}, Jun. 2016.

\bibitem{DBLP:GregorDGW15}
\BIBentryALTinterwordspacing
A.~G. D.~W. Karol~Gregor, Ivo~Danihelka, ``{DRAW:} {A} recurrent neural network
  for image generation,'' \emph{arXiv}, Feb. 2015. [Online]. Available:
  \url{https://arxiv.org/abs/1502.04623}
\BIBentrySTDinterwordspacing

\bibitem{introvae}
H.~Huang, R.~He, Z.~Sun, T.~Tan \emph{et~al.}, ``Introvae: Introspective
  variational autoencoders for photographic image synthesis,'' in \emph{NIPS
  2018}, Montreal, Canada, Nov. 2018, pp. 52--63.

\bibitem{ePC2017}
I.~Had\v{z}i\'{c}, Y.~Abe, and H.~C. Woithe, ``Edge computing in the e{PC}: A
  reality check,'' in \emph{Proceedings of the Second ACM/IEEE Symposium on
  Edge Computing}, San Jose/Fremont, CA, Oct. 2017.

\end{thebibliography}
\end{document}